\documentclass[english,aps, achemso,reprint,twocolumn,floatfix]{revtex4-1}

\usepackage[T1]{fontenc} 
\usepackage[latin9]{inputenc} 
\usepackage{graphicx}
\usepackage{float} 
\usepackage{amsmath}
\usepackage{textcomp}
\usepackage{gensymb}
\usepackage{amssymb}
\usepackage{tabularx}
\usepackage{amstext}
\usepackage{siunitx}
\usepackage{commath}
\usepackage{xcolor}
\usepackage{overpic} 
\usepackage{url}
\usepackage{hyperref}
\usepackage[normalem]{ulem}
\usepackage{colortbl}
\makeatletter

    \def\CT@@do@color{%
      \global\let\CT@do@color\relax
            \@tempdima\wd\z@
            \advance\@tempdima\@tempdimb
            \advance\@tempdima\@tempdimc
    \advance\@tempdimb\tabcolsep
    \advance\@tempdimc\tabcolsep
    \advance\@tempdima2\tabcolsep
            \kern-\@tempdimb
            \leaders\vrule
                    \hskip\@tempdima\@plus  1fill
            \kern-\@tempdimc
            \hskip-\wd\z@ \@plus -1fill }
    \makeatother
\usepackage{multirow}
\usepackage{breakcites}
\usepackage[final]{microtype}
\begin{document}

\title{
 Data-Driven Modeling of
 \texorpdfstring{${\rm S}_0 \rightarrow {\rm S}_1$}{S0 to S1}  Excitation Energy in 
 the BODIPY Chemical Space: 
 High-Throughput Computation,
 Quantum Machine Learning, and
 Inverse Design 
}

\date{\today}     

\author{Amit Gupta$^1$}
\author{Sabyasachi Chakraborty$^1$}
\author{Debashree Ghosh$^2$}
\author{Raghunathan Ramakrishnan$^1$}
\email{ramakrishnan@tifrh.res.in}

\affiliation{$^1$Tata Institute of Fundamental Research, Centre for Interdisciplinary Sciences, Hyderabad 500107, India}
\affiliation{$^2$Indian Association for the Cultivation of Science, Kolkata 700032, India}

\begin{abstract}
Derivatives of BODIPY are popular fluorophores due to their 
synthetic feasibility, structural rigidity, high quantum yield, and tunable
spectroscopic properties. While the characteristic absorption maximum of
BODIPY is at 2.5 eV, combinations of functional groups and substitution sites can 
shift the peak position by $\pm1$ eV. Time-dependent long-range corrected hybrid density functional methods can model the lowest 
excitation energies offering a semi-quantitative precision of $\pm$0.3 eV. 
Alas, the chemical space of BODIPYs stemming from combinatorial 
introduction of---even a few dozen---substituents is too large for 
brute-force high-throughput modeling. To navigate this vast space, we select
77,412 molecules and train a kernel-based quantum machine learning model providing
$<2$\% hold-out error. 
Further reuse of the results presented here to navigate the entire BODIPY universe comprising over 253 giga ($253 \times 10^9$) 
molecules is demonstrated by inverse-designing candidates with desired target excitation energies.
\end{abstract}

\maketitle   

\section{Introduction}


Among small molecule fluorophores, BODIPYs (derivatives of BODIPY, 
4,4-difluoro-4-bora-3a,4a-diaza-s-indacene)
hold a centre-stage in chemical physics 
due to their high quantum yield, high
molar absorption coefficients, bleaching resistance, 
narrow emission spectra, and
low-toxicity\cite{zhang2013bodipy,kowada2015bodipy}. 
They can be tuned to
fluoresce from blue to near-infra-red regions of the 
solar spectrum by structural modifications\cite{gomez20108,umezawa2008bright,buyukcakir2009tetrastyryl}.
BODIPYs are used in a multitude of applications such as 
theranostics\cite{qi2020fine}, 
laser dyes\cite{ulrich2008chemistry,ortiz2010red}, 
electro-luminescent films\cite{lai2003electrogenerated}, 
light-harvesting arrays\cite{eccik2017light,yilmaz2006light,ziessel2013artificial}, ion-sensors\cite{ozcan2018fluorescent,bakthavatsalam2015tuning}, supra-molecular gels\cite{cherumukkil2018self}, photo-sensitizers\cite{swamy2020near,filatov2020heavy,awuah2012boron}, fluorescent stains\cite{dahim2002physical}, chemical sensors\cite{matsumoto2007thiol}, energy transfer cassettes\cite{avellanal2021photosensitizers,ueno2011encapsulated}, 
band gap modulation\cite{wu2020strategic}, photo-dynamic therapy\cite{kamkaew2013bodipy}, and solar cells\cite{singh2014evolution,kolemen2011optimization,erten2008panchromatic}. Further, the synthetic ease 
of accessing BODIPYs has allowed development of dual emissive compounds with conformation-specific excitation characteristics\cite{mukherjee2013dual,mukherjee2014fine,nandi2021effect}.

    \begin{figure*}[!htbp]
        \centering
        \includegraphics[width=\linewidth]{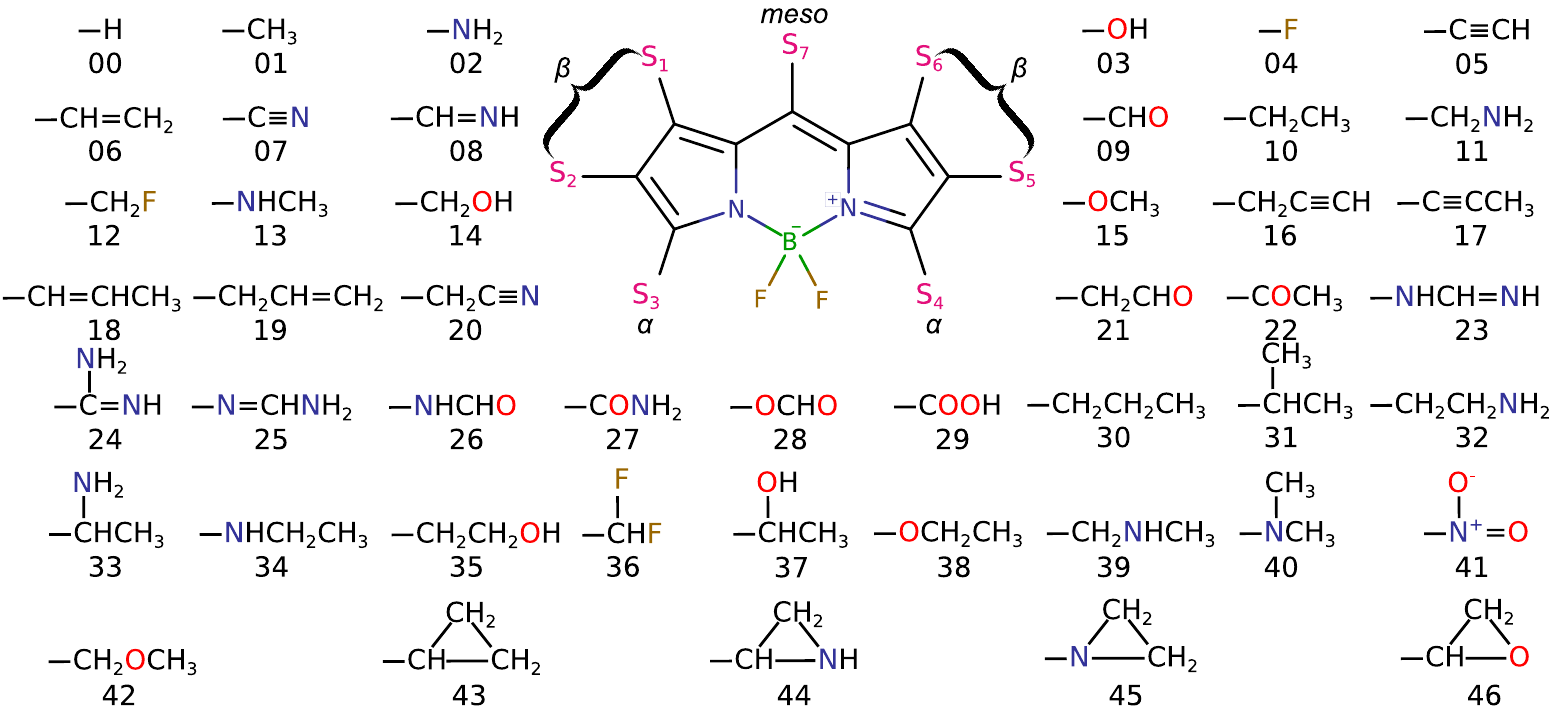}
        \caption{Composition of the BODIPY chemical space studied in this work. In the parent BODIPY molecule 
        all 7 carbon sites, S$_1$--S$_7$, are with H. Derivatives are obtained by combinatorially replacing H
        with 46 substituents.
        } 
        \label{fig:dataset}
    \end{figure*}

Even though the earliest report on their synthesis 
dates as far back as 1968\cite{treibs1968difluorboryl}, systematic explorations of BODIPYs became popular only in the 1990's\cite{banuelos2016bodipy,ulrich2008chemistry}.
However, the synthesis of unsubstituted BODIPY is relatively recent\cite{arroyo2009smallest,schmitt2009synthesis}. 
Advances in reaction methodologies and regioselective synthesis 
protocols have enabled targeted design of BODIPYs\cite{shimada2020regioselectively,feng2016regioselective,chen2009functionalization,wang2013uv}. Systematic
chemical mutations of BODIPY and their effects on the electronic spectra
 provided insights towards rational compound design\cite{donnelly2020exploring,tao2019tuning,lu2014structural}. 
As for their photochemical properties, 
substitution at the \textit{meso} position was found to offer
significant control\cite{donnelly2020exploring,prlj2016rationalizing}. 
While this collective experimental knowledge on BODIPYs represents the ground reality of their
electronic structure, there are known gaps in the chemical trends stemming from 
chemists' bias and synthetic limitations. In the case of organic photovoltaic materials,
high-throughput quantum chemistry combined with artificial intelligence algorithms has provided a bias-free solution for
better characterization\cite{hachmann2011harvard}.  

Among quantum chemistry approaches for excited state modeling, those revered for 
their favorable speed includes 
equation of motion coupled cluster with
singles and doubles, EOM-CCSD,
approximate CCSD, CC2\cite{christiansen1995second}, and
algebraic diagrammatic construction method in second-order perturbation theory\cite{dreuw2015algebraic}. Techniques like
spin-component-scaling and scaled-opposite-spin 
improve all three approaches\cite{tajti2019accuracy}. Their approximated versions
such as 
resolution-of-identity CC2, RI-CC2\cite{hattig2000cc2,send2011assessing}, and local pseudo natural orbital similarity transformed EOM-CCSD, DLPNO-STEOM-CCSD\cite{berraud2020unveiling}, 
retain the accuracy while
decreasing their computational scaling by an order. 
RI-CC2 has been used previously for
generating excited state spectra of a 
chemical space dataset with 22,780 small organic molecules\cite{ramakrishnan2015electronic}. 
For larger molecules, especially for high-throughput Big Data generation, 
the scaling offered by the aforementioned wavefunction 
methods are still unfavorable 
rendering time-dependent density 
functional theory, TDDFT \cite{gross2012introduction,runge1984density} as the preferred choice. 

Quantum machine learning (QML) methods\cite{rupp2012fast,ramakrishnan2017machine,von2018quantum} have come a long way 
from being tools for data analysis to be regarded as the `catalyst' in quantum chemistry big data campaigns\cite{hachmann2011harvard,ramakrishnan2015big,lopez2016harvard,beard2019comparative,abreha2019virtual}.
State-of-the-art structural descriptors facilitate inductive modeling 
of ground state properties with prediction accuracies better than that of
modern DFT approximations\cite{ramakrishnan2017machine,goscinski2021role}. For excited state
properties, in general, the error rates in QML 
have been noted to be inferior
compared to that of ground state properties\cite{faber2018alchemical,ye2020predicting,liu2021transferable,mazouin2021selected}. 
Yet, QML methods continue to find applications in excited state modeling 
in chemical space datasets\cite{ramakrishnan2015electronic,tapavicza2021elucidating,caylak2019evolutionary,caylak2021machine} as well as in
potential surface manifolds\cite{westermayr2020deep,hase2020designing,koerstz2021high,ju2021machine,kiyohara2020learning,westermayr2020machine}. 
Keeping abreast with the progress in QML,
materials/molecules inverse-design protocols have also advanced
since the earliest implementation nearly twenty years ago\cite{franceschetti1999inverse}. 
Wang \textit{et al.}\cite{wang2013accurate} employed the extreme ML neural network with descriptors of varying rigour to predict experimental excitation properties of selected BODIPYs. Huwig \textit{et al.}\cite{huwig2017properties} inverse-designed benzene derivatives with different properties preferred for dye-sensitized solar cell applications. Recently, Lu \textit{et al.}\cite{lu2021accelerated} inverse designed BODIPY dyes, for applications in dye-sensitized solar cells.

In the present study, we enumerate the complete chemical space formed 
by combinatorially introducing 46 small organic groups with up to 3 CONF atoms
at all free sites of BODIPY. We generate geometries of all possible singly and doubly substituted BODIPYs, and randomly drawn subsets with triple-to-septuple substitutions. For the resulting set of 77,412 molecules, we obtain accurate
DFT-level geometries and determine excited state properties with TDDFT. 
We perform detailed chemical analyses on the 
functional group modulation of the
lowest excitation energy corresponding to the brightest state.
The dataset generated is used to benchmark the performance of a
kernel-based QML approaches for modeling excitation energies
with various molecular descriptors. 
Using the best model as a property generator, we 
embark on inverse-designing BODIPY molecules with target excitation energies.



\section{Data and Methods} \label{dataset}

\subsection{BODIPYs Chemical Space Design}
The size of the BODIPY chemical space formed by combinatorially introducing functional groups at all the free sites is countably infinite. A suitable molecular subspace may be identified by limiting the size of these
functional groups. 
While BODIPYs with varied Stokes shifts for multicolor fluorescence microscopy have been developed\cite{bittel2018varied}, 
most exhibit only modest shifts 
suggesting the excited-state geometries to be very similar 
to that in the ground-state\cite{loudet2007bodipy,ziessel2007chemistry}. 
Hence, it is sufficient to model only the vertical excitation energies of BODIPYs without adiabatic considerations.
To this end, we select a set of 46 small organic substituents and combinatorially introduce them
at the 7 free sites of BODIPY 
(two $\alpha$: sites-3,4, 
four $\beta$: sites-1,6, sites-2,5, 
and one: \textit{meso}: site-7), 
as presented in FIG.~\ref{fig:dataset}.
Further, we explore only those derivatives formed by single-bond connectivities and avoid substituents leading to fused rings. 
To keep the substituents devoid of chemists' bias, we sampled them from the smallest molecules of the QM9 database\cite{ramakrishnan2014quantum}. Some functional
groups missing in the QM9 molecules have been introduced for the 
sake of completeness providing groups spanning a spectrum of
 electron-donating/electron-withdrawing capacity. 
 \begin{table}[!htbp]
    \centering
    \caption{Size of the BODIPY chemical space considered in this study. The 
    numbers correspond to unique molecules formed by replacing out of 7 H atoms
    in BODIPY with $N_s$ 
    substituents listed in FIG.~\ref{fig:dataset}. 
    The enumeration was performed algebraically. Numbers in parentheses 
    correspond to molecules for which 
    coordinates were generated for DFT and QML modeling.}
    \begin{tabular}{l c r r}
    \hline
    \multicolumn{1}{l}{$N_s$} &  \multicolumn{1}{c}{} & \multicolumn{1}{l}{Unique molecules}  & \\
    \hline
    0 &  & 1&(1) \\
    1 &  & 184& (184) \\
    2 &  & 22,287& (22,287)\\
    3 &  & 1,706,554 &(10,999)\\ 
    4 &  & 78,358,654 &(10,990)\\
    5 &  & 2,162,757,252 &(10,982)\\
    6 &  & 33,160,087,804 &(10,986)\\
    7 &  & 217,911,067,336 &(10,983)\\
    \hline
    \multicolumn{1}{l}{Total} & & 253,314,000,072 &(77,412)\\
    \hline
    \end{tabular}
    \label{tab:enum}
\end{table}
%
 
 
On an asymmetric framework, the total number of molecules that can be 
formed by introducing 46 groups in 7 sites should be $46^7=435.8\times10^9$. However,
since the unsubstituted BODIPY framework has the C$_{2v}$ point group symmetry\cite{pogonin2020quantum,teets2009three}, this number drops 
when redundant entries are eliminated. For such symmetry constrained enumerations, P{\'o}lya\cite{polya1937kombinatorische,freudenstein1967basic,polya2012combinatorial} has suggested an algebraic strategy that has been used for non-constructive enumeration of chemical compound spaces\cite{chakraborty2019chemical}. 
With in the constraints of C$_{2v}$, the total number of molecules in the
BODIPYs chemical space considered here amounts to $253\times10^9$, as 
reported in Table.~\ref{tab:enum}. We 
selected all compounds with up to 2 substitutions (22,472 molecules) 
and 11,000 entries with 3--7 substitutions. 
In the latter categories 60 molecules exhibited poor
convergence in DFT calculations and  were discarded. The resulting set of
77,412 unique BODIPY molecules were utilized for training QML models. 


\subsection{Quantum Chemistry}\label{sec:DFT}
Initial geometries of 77,412 BODIPY molecules were generated by the ``lego approach''---three dimensional
structures of substituents attached to the BODIPY sites. These geometries were relaxed using the universal
force field (UFF)\cite{rappe1992uff} as implemented in Openbabel\cite{o2011open}. Subsequently, these geometries 
were relaxed with the semi-empirical method, PM7\cite{stewart2013optimization} 
available in MOPAC2016\cite{MOPAC2016}. Finally, the geometries were optimized to
their minimum energy configurations at the B3LYP level\cite{becke1993becke} with the Weigend basis set, def2-SVP.
B3LYP calcuations were accelerated with RI\cite{vahtras1993integral,kendall1997impact} 
using the Weigend auxiliary basis sets\cite{weigend2006accurate}, as implemented in 
TURBOMOLE\cite{furche2014turbomole}.
At the TDDFT level, CAM-B3LYP\cite{yanai2004new}/def2-TZVP, we calculated the lowest ten excited states of all 77,412 molecules in a single-points fashion using the B3LYP/def2-SVP geometries. TDDFT calculations were accelerated
by RIJCOSX, RI approximation for Coulomb (J) and `Chain-Of-Spheres' (COS) algorithm for exchange integrals, as implemented in ORCA\cite{neese2012orca}. 

The performance of long-range corrected hybrid DFT functionals for excited spectra
is well-established for diverse benchmark datasets\cite{goerigk2010assessment}. 
Amongst these, CAM-B3LYP\cite{yanai2004new} presents good correlations with experimental\cite{momeni2015td}, EOM-CCSD\cite{bose2016effect}, and CC2 results\cite{shao2019benchmarking}. Thus, for a selected subset of BODIPY derivatives, 
we benchmarked CAM-B3LYP's accuracy along with that of BLYP\cite{becke1988density}, B3LYP\cite{becke1993becke} against 
STEOM-DLPNO-CCSD method\cite{berraud2020unveiling} with the def2-TZVP basis set using ORCA. 
 Additionally, CAM-B3LYP's performance is also tested against the
 SOS-CIS(D) method\cite{grimme2004calculation,rhee2007scaled} with aug-cc-pVDZ basis set
 using QCHEM\cite{shao2015advances}.
 The latter wavefunction method is known to exhibit good accuracy
 for excitation energies\cite{chibani2014improving,bose2017electrostatic} with experimental results. Here, we want to 
 compare its performance with STEOM-DLPNO-CCSD for modeling BODIPY's ${\rm S}_0 \rightarrow {\rm S}_1$ excitation energy. 



\subsection{Machine Learning}
In the present study, we employed the kernel 
ridge regression (KRR) QML method 
for its accuracy, scalability, and interpretability marked by successes in various endeavors\cite{huang2020quantum,gupta2021revving,kramer2020charge,faber2018alchemical,huang2020quantumnadine,rupp2015machine}. 
For various molecular properties, the learning rates of KRR-QML was shown to improve with increasing training-set size\cite{ramakrishnan2015many,huang2020quantum,gupta2021revving,kramer2020charge,tapavicza2021elucidating,faber2018alchemical,huang2020quantumnadine}. In KRR, property modeling is posed as a regression problem using the `kernel trick', where a higher dimensional feature space is sampled using a kernel function. 
Hence, the regression problem can be expressed as 
$\left( {\bf K} + \lambda {\bf I} \right) {\bf c} = {\bf p}$, 
where ${\bf K}$ is the kernel matrix, 
$\lambda$ is a hyperparameter quantifying the regularization strength, 
${\bf c}$ is the regression coefficient vector,
and ${\bf p}$ is the target property vector. 
The elements of the positive-semi-definite kernel matrix are given by 
$k_{ij} = k({\bf d}_i,{\bf d}_j) \in (0,1]$, where ${\bf d}_i$ is the descriptor
vector for the $i$-th entry. For the choice of descriptor, we benchmarked the performances of a 
1-hot representation along with the structural descriptors:
Bag-of-Bonds\cite{hansen2015machine}(BoB), 
Felix-Christensen-Huang-Lilienfeld (FCHL)\cite{faber2018alchemical}, and 
Spectrum of London and Axilrod-Teller-Muto potential (SLATM)\cite{huang2020quantum}. SLATM and FCHL descriptors were generated using the QML package\cite{christensen2017qml}, while BoB and 1-hot vector using an in-house code.
The 1-hot representation was shown to perform well when the dataset is combinatorially
diverse\cite{faber2016machine,ward2017including,kayastha2021machine,heinen2021toward}.
The 1-hot representation is a 322-bit ($7 \times 46$) vector, where the presence/absence of 
one of the 46 substituents at the 7 sites is denoted by 1/0. This representation does not include any 
information on the BODIPY framework that is common to all molecules.



For the choice of kernel function,
we used the Laplacian function, $k({\bf d}_i,{\bf d}_j)= \exp(- |{\bf d}_i - {\bf d}_j|_1 / \sigma )$, where $|\cdot|_1$ denotes the L$_1$ norm while $\sigma$ is the hyperparameter quantifying kernel width. 
For FCHL, we determined an optimal kernel width of $\sigma=5$ by scanning with a fixed regularization strength along with a cutoff of 5 \AA. 
For selecting hyperparameters,
we followed the single-kernel strategy\cite{ramakrishnan2015many}. When there is 
no linear dependency in the reproducing kernel Hilbert space\cite{ho2003reproducing}, $\lambda$ can be exactly set
to 0.0. To prevent near linear dependency rendering the kernel matrix 
singular due to finite precision, especially for large training sets, we used a small
value of $\lambda=0.001$ throughout, as in our previous work on ML modeling of ${}^{13}$C NMR
shielding constants\cite{gupta2021revving}. The kernel width was selected using 
the sample median of all descriptor differences, $d_{ij}^{\rm median}={\rm median}\{|{\bf d}_i - {\bf d}_j|_1 \}$, as 
$\sigma=d^{\rm median}_{ij}/(\log 2)$\cite{ramakrishnan2015many}. For 1-hot/BoB/SLATM representations,
the optimal $\sigma$ was found to be 26.57/3603.98/840.09.
All structural representations were calculated using PM7 level minimum energy geometries
to facilitate rapid querying in the BODIPY chemical space with QML. 

\subsection{Machine Learning aided Inverse-Design}
For inverse-design of BODIPY derivatives with a desired S$_0$ $\rightarrow$ S$_1$ excitation
energy, we used a trained QML model as a rapid surrogate to DFT. 
We found the QML model based on the SLATM descriptor to deliver
best learning rates as discussed later in Section~\ref{sec:MLbenchmark}.
For minimizing molecular configuration variables in the property manifold defined by the QML model, 
we explored Bayesian optimization and genetic algorithm (GA)
operating in the SLATM feature space.

\subsubsection{Bayesian optimization}
The Bayesian optimization method is self-correcting\cite{jones1998efficient}. Its performance improves over iterations by using previously
sampled attribute-value (i.e. descriptor-property) pairs as prior. The `gradient' for sampling the next entry is estimated by Gaussian process regression\cite{rasmussen2003gaussian}. The process begins with a normally distributed ($\mathcal{N}$) 
sample space of descriptor vectors of a training set, ${\bf t}=\{ t_1,t_2,\ldots\}$,
along with the corresponding property values, ${\bf p}_{\bf t}=\{ p_{t_1},p_{t_2},\ldots\}$. 
\begin{equation}
    {\bf p}_{\rm t} \thicksim \mathcal{N}({\bf p}_{\rm t}, \boldsymbol{\Sigma}_{\bf t t}),
\end{equation}
where $\boldsymbol{\Sigma}_{\bf t t}$ is the positive definite covariance matrix, taken as the Gaussian kernel matrix with an added noise.  
For a set of query molecules, ${\bf q}=\{q_1,q_2,\ldots \}$, the target property values
and their uncertainties are predicted as the mean and variance of a Normal distribution,
${\bf p}_{\bf q} \thicksim \mathcal{N}(\boldsymbol{\mu_*},\boldsymbol{\Sigma_*})$. The
estimated mean values, $\boldsymbol{\mu_*}$ and variances, $\boldsymbol{\Sigma_*}$, are given as
 \begin{align}
    {\bf p}_{\bf q}\thicksim\boldsymbol{\mu_*}=\left[ \boldsymbol{\Sigma}_{\bf t q} \right]^{\rm T}\left[ \boldsymbol{\Sigma}_{\bf t t}\right]^{-1}{\bf p}_{\bf t}.
\end{align}
Prediction variance is given by the diagonal elements of the matrix
\begin{align}
    \boldsymbol{\Sigma_*} = 
     \boldsymbol{\Sigma} _{\bf q q}- 
    \left[ \boldsymbol{\Sigma}_{\bf t q} \right]^{\rm T}
    \left[ \boldsymbol{\Sigma} _{\bf t t}\right]^{-1} 
     \boldsymbol{\Sigma}_{\bf t q}. 
\end{align}
Hence, the predicted property for a query is the $q$-th element of $\boldsymbol{\mu_*}$, $\mu_q$.
The corresponding variance, $\sigma_q$, is the diagonal element of $\boldsymbol{\Sigma} _{\bf q q}$ at row-$q$ and column-$q$.
New sampling points are proposed using an acquisition function, $\mathcal{A}$. A popular choice for $\mathcal{A}$ is the expected improvement defined as 
\begin{equation}
    \mathcal{A}(q) = 
    \begin{cases}
        Z_q \Phi(-\frac{Z_q}{\sigma_q}) + 
        \sigma_q \phi(-\frac{Z_q}{\sigma_q}), & \text{if } \sigma_q > 0\\
        0, & \text{if } \sigma_q = 0
    \end{cases}
\end{equation}
where $Z_q = \mu_q - {\rm max}(\{ \boldsymbol{\mu_*} \}) + \zeta$, the set
$\{ \boldsymbol{\mu_*} \}$ contains all values 
sampled until a given iteration,
and $\zeta$ is a real-valued hyper-parameter, while $\Phi$ and
$\phi$ are the cumulative and probability distribution functions of $\mathcal{N}$, respectively. 

\subsubsection{Genetic Algorithm}
Genetic algorithm (GA) is an evolution-inspired heuristic method 
for optimization in a high-dimensional space with combinatorially coupled variables\cite{whitley1994genetic}. 
For sampling in chemical space, GA has been 
shown to be a suitable framework\cite{browning2017genetic,verhellen2020illuminating}.
In this study, we initialized 
the first generation in GA optimization 
with a population of 20 random molecules, and a mutation rate of 0.01. 
Additionally, in each generation, we populated the sample with 
10 random molecules. For the entire sample, 
$E ({\rm S}_0 \rightarrow {\rm S}_1)$ was predicted by an ML model
trained on DFT-level properties. Absolute deviation of these energies from the
target value was used as the fitness, and only molecules with deviations
smaller than the population median entered subsequent generations
through crossover. 
Our implementation of the Bayesian and GA optimizations
along with sample input files and details of control parameters
are available at \href{https://github.com/moldis-group/bodipy}{https://github.com/moldis-group/bodipy}.

\section{Results and Discussion}

\subsection{\texorpdfstring{Chemical trends in ${\rm S}_0 \rightarrow {\rm S}_1$ excitation energy }{Chemical trends in S0 to S1}}
Wavelength tuning of BODIPY by controlled synthesis has been successful 
for a handful of symmetrically substituted derivatives\cite{lu2014structural}. 
 \begin{figure}[!htp]
    \centering
    \includegraphics[width=\linewidth]{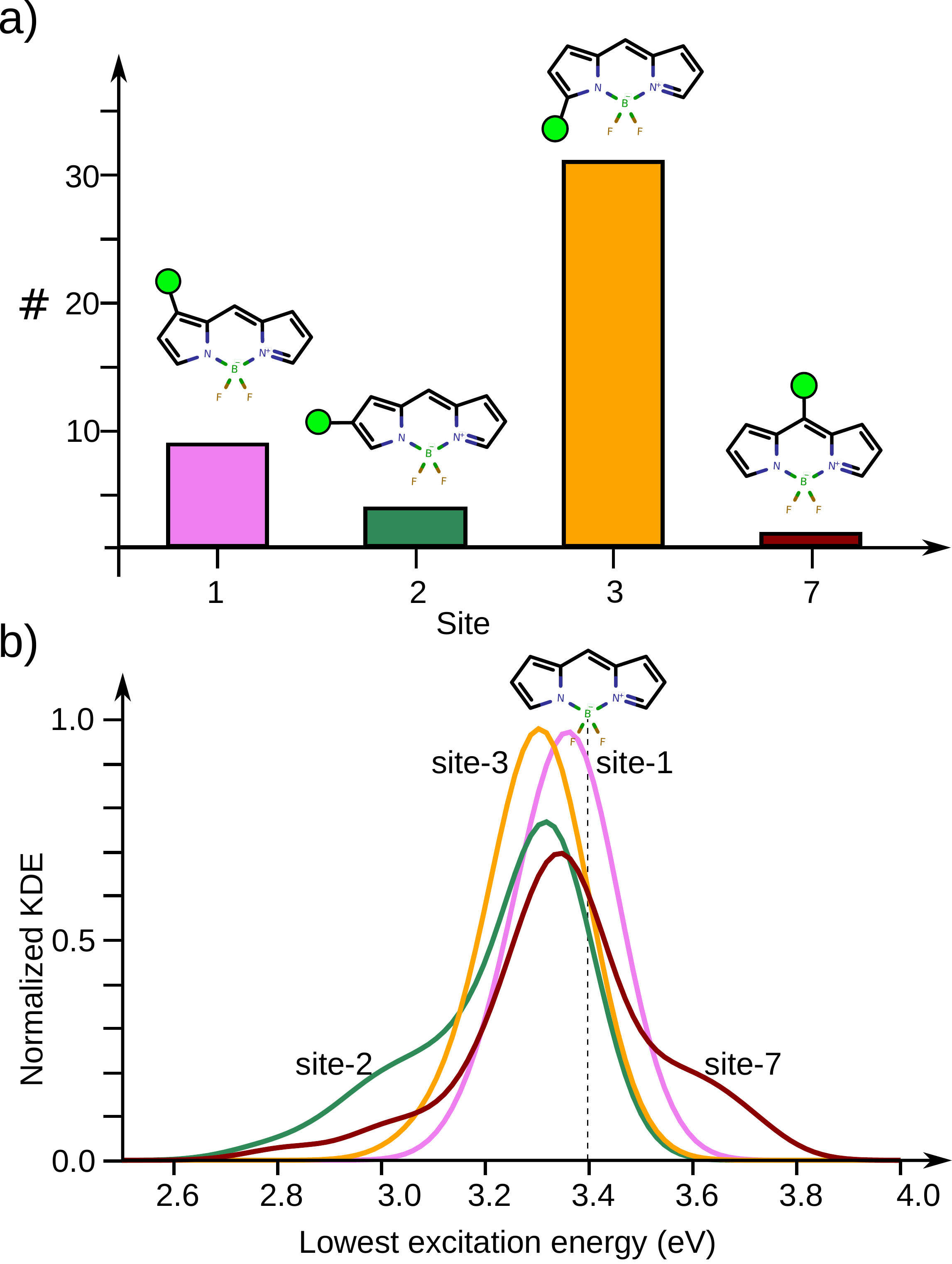}
    \caption{Site-specific statistics for singly-substituted BODIPYs.
    a) Frequencies of thermodynamically most preferred site for 46 substituents.
    b) Distribution of S$_0\rightarrow$S$_1$ excitation energies for 46 singly-substituted BODIPY molecules per site. Vertical dashed line marks the value for unsubstituted BODIPY.
    }
    \label{fig:1DDFT}
 \end{figure}
 A more comprehensive picture of the dependence of the wavelength shift, 
corresponding to the brightest excitation, on chemical 
factors requires further evidences sampled across
a larger chemical space. Herein, we investigate the roles 
played by the substitution sites and the groups in modulating
BODIPY's stability and excitation characteristics. 

To identify a suitable level of DFT approximation,
for high-throughput modeling, we benchmarked the ${\rm S}_0 \rightarrow {\rm S}_1$ excitation
energies of the unsubstituted BODIPY and 184 of its singly-substituted derivatives. For references, we
used STEOM-DLPNO-CCSD/def2-TZVP and SOS-CIS(D)/aug-cc-pVDZ results. Five singly substituted derivatives 
failed to converge at the reference wavefunction-level calculations and these were not included for benchmarking.
Compared to STEOM-DLPNO-CCSD/def2-TZVP we obtained mean absolute errors (MAEs) of 
0.31$\pm$0.33/0.13$\pm$0.16/0.05$\pm$0.06 eV for 
BLYP/B3LYP/CAM-B3LYP DFT methods with the def2-TZVP basis set.  
CAM-B3LYP values were also found to agree with the
SOS-CIS(D)/aug-cc-pVDZ level yielding an MAE of 0.05$\pm$0.05 eV. Hence, we performed all TDDFT calculations at the 
CAM-B3LYP/def2-TZVP level. Even though the SOS-CIS(D) and the STEOM-DLPNO-CCSD calculations have been performed 
with different basis sets, they agree well with a coefficient-of-correlation ($R^2$) of 0.82 and an average
deviation of 0.04$\pm$0.05 eV. Hence, we conclude the residual errors in CAM-B3LYP/def2-TZVP based
excited state results of the BODIPYs dataset presented here to be with in the uncertainties 
expected across wavefunction methods and basis set definitions.

\begin{figure*}[!htp]
        \centering
        \includegraphics[width=\linewidth]{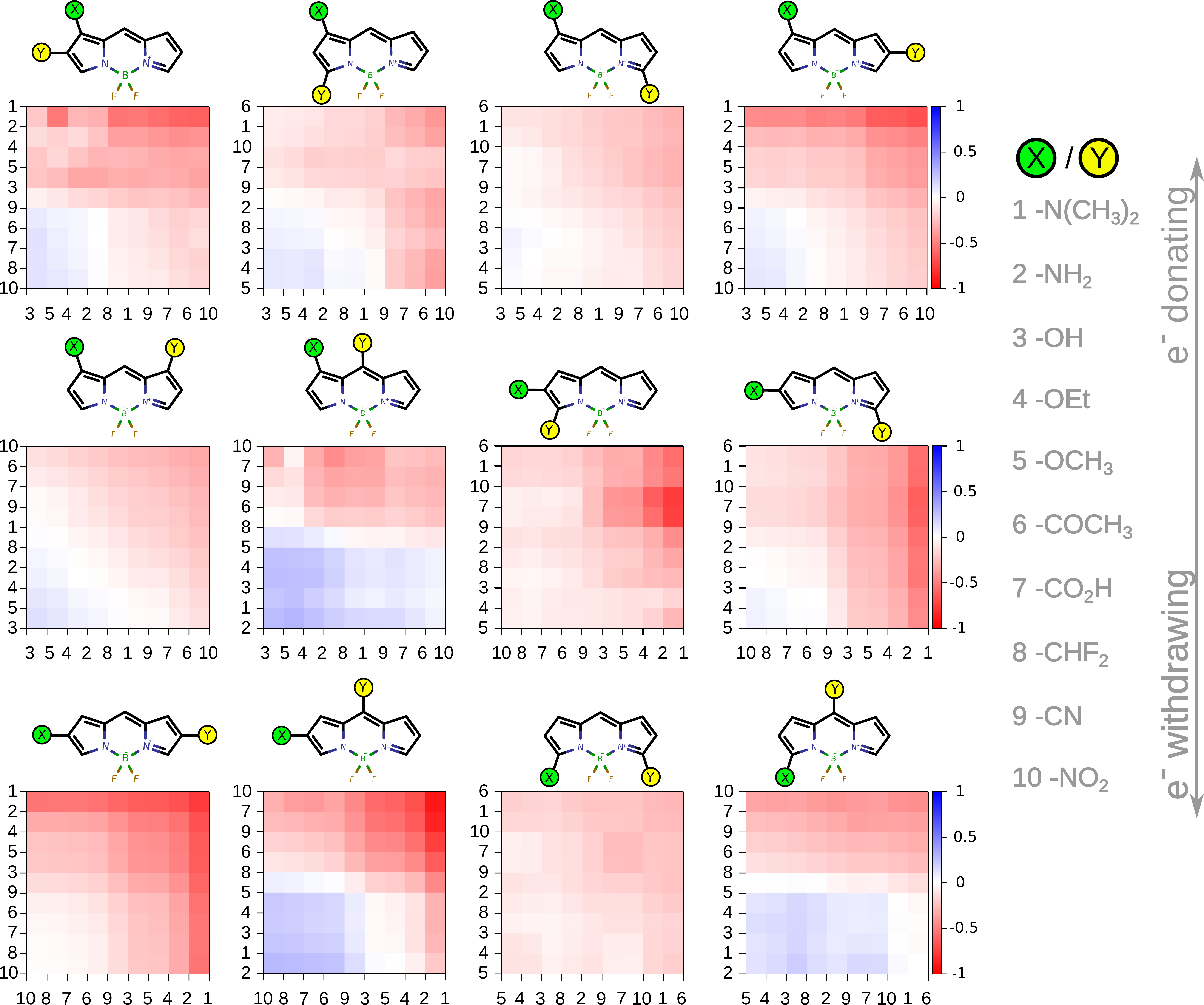}
        \caption{Modulation of  
        S$_0\rightarrow$S$_1$ excitation energy with substitution.
        For 12 unique double substitution patterns, shifts in 
        excitation energy, $\Delta E=E{\rm (substituted)}-E{\rm(BODIPY)}$,
        are presented. For clarity, only 10 substituents that are  
        a standard scale of electron-donating/-withdrawing 
        are drawn from the 46-set shown in FIG~\ref{fig:dataset}. 
        In each of the 12 panels, X and Y ordering is according to
        shifts in singly-substituted compounds. 
        }
        \label{fig:dataset3}
\end{figure*}
 \begin{table}[!hbp]
    \centering
    \caption{Error metrics for a group-additive estimation of 
    the lowest excitation energy with respect to actual TDDFT values for multiply substituted BODIPY molecules: 
    mean absolute error (MAE in eV), 
    standard deviation (SD in eV),
    mean percentage absolute error (MPAE), and
    coefficient of determination ($R^2$) 
    are presented.}
    \begin{tabular}{p{2.0cm} p{1.0cm} p{1.0cm} p{1.0cm} p{1.0cm} p{1.0cm} }
    \hline
    \multicolumn{1}{l}{Substitution}     &  
    \multicolumn{1}{l}{MAE}  & 
    \multicolumn{1}{l}{SD} &
    \multicolumn{1}{l}{MPAE} &
    \multicolumn{1}{l}{$R^2$}  
    \\
    \hline
    doubles       & $~$0.03 & 0.06  & 1.07 & $~$0.87    \\
    triples       & $~$0.06 & 0.09  & 2.09 & $~$0.76   \\
    quadruples    & $~$0.10 & 0.13  & 3.33 & $~$0.62    \\
    quintuples    & $~$0.14 & 0.15  & 4.69 & $~$0.49   \\
    hextuples     & $~$0.19 & 0.19  & 6.38 & $~$0.20   \\
    septuples     & $~$0.24 & 0.21  & 8.12 & -0.01    \\
    \hline
    \end{tabular}
    \label{tab:devstats}
\end{table}

In FIG.~\ref{fig:1DDFT}a, we present the frequencies of the thermodynamically favored site for all 46 substituents. 
This qualitative trend will be reflected in their relative 
synthetic yields at the high-temperature limit. 
We find site-3 to be preferred by 
most groups, as previously noted for alkyl-based substituents\cite{mukherjee2015effect},
followed
by site-1. Accommodating the substituents in both sites (1 \& 3) result in minimal
perturbation of the TDDFT excitation energy of
 BODIPY scaffold at 3.40 eV, see FIG.~\ref{fig:1DDFT}b. This value deviates by 0.91 eV from the more reliable
 STEOM-DLPNO-CCSD value of 2.49 eV. The latter, is in excellent agreement with the experimental value 
 $\lambda_{\rm max}^{\rm abs.}=503$ nm (2.46 eV) and  $\lambda_{\rm max}^{\rm em.}=512$ nm (2.42 eV)\cite{schmitt2009synthesis}.

A small systematic 
red shift of the site-3 values maybe ascribed to the non-bonding interactions between
the groups and an F atom of BODIPY. Site-2  ($\beta$) and site-7 ({\it meso}), 
that are thermodynamically least preferred 
also result in strong shifts of the S$_0\rightarrow$S$_1$ excitation energy,  FIG.~\ref{fig:1DDFT}b. 
Of particular interest, substitutions at site-2 mostly red-shifts the base excitation energy of BODIPY 
while that at site-7 results in blue-shifting.
Chemical non-equivalence of site-2 compared to the other sites has been
rationalized by the presence of a node in the lowest unoccupied 
molecular orbital (LUMO)\cite{lu2014structural}.  
The most blue-shifting substituent/site combination corresponds to ethylamine at site-7 (excitation energy at 3.70 eV), while the most red-shifting one is dimethylamine at site-2 (excitation energy at 2.78 eV). It is interesting to note that both 
ethylamine and dimethylamine are groups with similar electron-donating capacity. 
Hence, at least
for the case of singly substituted BODIPY derivatives, thermodynamic stabilities and the 
shift of excitation energy are largely controlled by the substitution site.

The singly substituted BODIPY derivatives show substitution on site 7 to provide the most versatile tuning followed by sites 2/5, 3/4, and 1/6. However, it does not disclose `inter-substituent' interactions affecting the overall excitation properties. To gauge the exact nature and extent of these inter-substituent interactions, we inspect the deviation of the TDDFT value of $E_{\mathrm{S}_{0}\rightarrow \mathrm{S}_{1}}$ of an $n$-tuply substituted BODIPY derivative from that of values estimated by
employing additivity principle
\begin{equation}
\begin{aligned}
    E_{\mathrm{S}_{0}\rightarrow \mathrm{S}_{1}} = & E_{\mathrm{S}_{0}\rightarrow \mathrm{S}_{1}}^{\rm BODIPY} + 
    \Delta E_n (s_1^{g_1},\ldots,s_n^{g_n}).
\end{aligned}
\label{eq:add}
\end{equation}
Here, $E_{\mathrm{S}_{0}\rightarrow \mathrm{S}_{1}}$ 
is the first excitation energy
of a BODIPY derivative, 
$E_{\mathrm{S}_{0}\rightarrow \mathrm{S}_{1}}^{\rm BODIPY}$ being the value corresponding to the unsubstituted BODIPY.
For a singly-substituted derivative with
group $1\le g_1 \le 46$ at site $1 \le s_1 \le 7$,
an exact shift, $\Delta E_1(s_1^{g_1})$, is calculated as 
the difference between the singly-substituted and
unsubstituted BODIPYs.

For Eq.~\ref{eq:add} to be of practical use in estimating the 
energy of an arbitrary derivative, the higher-order 
corrections should be approximated by lower-order terms. Here, we use $\Delta E_{1}(s_1^{g_1})$ determined for the singly-substituted derivatives to approximate the higher-order terms as the sum
\begin{equation}
    \Delta E_n (s_1^{g_1},\ldots,s_n^{g_n}) \approx 
    \sum_{k=1}^n \Delta E_{1}(s_k^{g_k}).
    \label{eq:approx}
\end{equation}
In Table.~\ref{tab:devstats}, we present the statistics for the
estimation of TDDFT excitation energies of 77 k BODIPYs,
for exact counts
of molecules, see Table.~\ref{tab:enum}. The estimated values
of 22 k doubly substituted ($n=2$) derivatives 
show a good agreement with the
target TDDFT values with a mean absolute error (MAE) of 0.03 eV and a 
Spearman rank correlation ($\rho$) of 0.95, see Table~\ref{tab:devstats}.

The agreement between
the reference values and the additivity model diminishes with
increase in the number of substituents. For every additional substituent, the increase in error is 0.03--0.04 eV. 
A similar increase in the standard
deviation suggest the errors to have
non-systematic contributions. For the limiting case, $n=7$,
the prediction MAE is $>0.2$ eV---over $8\%$ of the reference 
values---which is comparable to the spread of excitation energy values.
Furthermore, the $R^2$ value for estimations was 
found to be essentially zero. 
Since a large fraction of the BODIPYs chemical space comprises
of septuply-substituted molecules, such large errors 
make group additive estimation a poor baseline 
for $\Delta-$QML\cite{ramakrishnan2015big}. While systematic
diagnostics to quantify a method as a 
baseline is still lacking, in our past $\Delta-$QML works, we found 
better learning rates when $R^2>0.5$ when comparing the baseline and targetline values. 
In $\Delta-$QML modeling of DFT-level
$^{13}$C NMR shielding of
QM9 molecules, a minimal basis set baseline yielded $R^2=0.66$ 
resulting in better learning rates than modeling directly on the
target values\cite{gupta2021revving}. 

Upon double substitution, we note
the range of S$_0\rightarrow$S$_1$ excitation to increase compared to the singly substituted compounds. While the majority of BODIPYs show red-shifted $\lambda_{\rm max}$, compared
to the unsubstituted molecule, combinatorial exploration with
high-throughput first-principles calculations identifies a non-negligible fraction of blue-shifted molecules. Out of 22,287 doubly-substituted BODIPYs studied here, 95.45\% corresponding to 21,272 entries have large S$_0\rightarrow$S$_1$ oscillator strengths ($f>0.5$) suggesting good potential for light-harvesting applications. 
However, only 1,884 of these candidates
were found to be blue-shifted. 
The increase in the approximation error in Eq.\ref{eq:approx}, 
with increasing $n$ may be expected as the joint chemical effects due to multiple substituents can no 
longer be treated as a weak perturbations. We inspect the
$n=2$ case corresponding to the excitation energies of 22,287 
doubly substituted BODIPYs to identify the combinations of sites/groups 
resulting in non-additive trends. For this purpose, we selected 10
representative substituents (out of 46) with well-characterized 
electron-donating and withdrawing abilities compared to the standard
aromatic molecule, benzene. 

FIG.~\ref{fig:dataset3} presents the shift in  $E( {\rm S}_0 \rightarrow {\rm S}_1)$ across 12 positional isomers of 
doubly substituted BODIPY derivatives with respect to unsubstituted BODIPY. The order of X and Y axes 
are independently sorted according to the shifts in the 
singly-substituted series. Hence, in FIG.~\ref{fig:dataset3}, for the first heatmap (top, left-most), at site 1 we have X substituents along the X-axis while the substituents at site 2 have Y substituents on Y-axis. In the
selected color scale, the blue-shifted molecules will appear blue while the red-shifted entries appear in red. For symmetric combinations of 
sites, where (X,Y) = (1,6), (2,5), and (3,4), 
the number of unique molecules is 55---for consistency, 
these heatmaps are
presented by duplicating entries. 
The group additivity model (Eq.~6) does not differentiate positional isomers. For instance, since site-2 and site-5 are equivalent for single substitution, for the combination (1,2) and (1,5) the additivity model will predict same shifts of the excitation energy. Since sites 1 and 5
are spatially separated, the additivity assumption holds better resulting in a smooth transition in the heatmap. However, for (1,2) double
substitutions, the effect of inter-substituent interaction is reflected in irregularities in the diagonal gradient (from left/bottom to right/top). Similar trend holds for the (2,3)-vs.-(2,4) case. Substitutions at sites 1 \& 3 result in mild shifts as seen in FIG.~\ref{fig:dataset3}. Hence, (1,4)-derivatives (with weak inter-substituent interactions) yield
weak shifts compared to isomerically equivalent (1,3) substitutions.



Blue-shifting of the absorption spectra due to \textit{meso} substitution (at site-7) was previously
observed\cite{banuelos2011new}. In FIG.~\ref{fig:1DDFT}, we see this effect for the singly-substituted derivatives. Hence, out of 12 unique doubly substituted patterns, blue-shifting is predominantly
noted for three combinations involving site-7: (1,7), (2,7), and (3,7). Of these, since substitutions at sites 1 \& 3 result
in weak perturbations of BODIPY's excitation characteristics, a larger fraction of blue-shifted derivatives are seen for
(1,7) and (3,7) substitutions. Since single substitutions at sites 2 \& 7 have shown strong but contrasting
trends, see FIG.~\ref{fig:1DDFT}, their joint occurrence shows constructive and destructive effects. 
Of the two weakly perturbing sites, 1 \& 3, the former results in a slightly blue-shifted distribution, while the latter in a slightly
red-shifted distribution, see FIG.~\ref{fig:1DDFT}b. The dependence of the shift with electron donating/withdrawing ability of substituents
is similar for sites 1, 3, \& 7. Hence, to blue-shift BODIPY, having electron donating groups at sites (1,7) is
ideal. Similarly, to red-shift BODIPY, having an electron donating group at 2, and an electron withdrawing group at 3 or 4 is ideal. In FIG.~\ref{fig:dataset3},
we see the (2,3) combination to benefit from inter-substituent interactions over the (2,4) combination. We tested the validity of these
trends by identifying the extreme doubly-substituted molecules by considering the entire set of 46 substituents. 
The excitation energy of the most blue-shifted derivative appears at 3.84 eV (--OH at site 1 and --NHCH$_3$ at site 7),
while the most red-shifted appears at 2.32 eV (--NHCH$_2$CH$_3$ at site 2 and --COCH$_3$ at site 3).

\subsection{Quantum Machine Learning Models}\label{sec:MLbenchmark}
The main objective of QML modeling is to provide an inexpensive inference approach to replace rigorous, first-principles modeling.
For reliable high-throughput screening in the BODIPY chemical space, one cannot depend on an additivity model based on chemical
effects imparted by site-specific individual substitutions on BODIPY. With increasing substitutions, the additivity 
model ceases to be even qualitatively accurate, see Table.~\ref{tab:devstats}. On the other hand,
QML models when sufficiently trained using a baseline geometry can facilitate rapid querying in the uncharted regions across the chemical space.
Hence, QML offers an opportunity to effortlessly navigate across the vast BODIPY chemical space with quantitative accuracy. 
As discussed before (see Sec.~\ref{sec:DFT}), of the
77,412 molecules for which DFT calculations were performed, QML modeling was done
with 76,212 entries in the training set. The unsubstituted, and all 184 singly-substituted
molecules were kept in training. Of the 22,287
doubly substituted derivatives, 200 was kept in a hold-out set.
\begin{table*}[!htbp]
    \caption{Breakdown of ML errors for predicting the lowest excitation
    energy of 1200 hold-out BODIPY derivatives. 
    Mean percentage absolute error (MPAE) is given separately for 
    200 entries with $N_s$ substituents. 
     }
        \begin{tabular}{r l llllll l llllll}
        \hline
        \multicolumn{1}{l}{$N$}  &  $~~~$& \multicolumn{13}{l}{MPAE for various $N_s$ }   \\
        \cline{3-15}
         &  & \multicolumn{6}{l}{SLATM} & $~~~$ & \multicolumn{6}{l}{1-hot}\\
        \cline{3-8}\cline{10-15}
        & & 2    & 3    & 4    &  5   & 6  &  7 &  & 2    & 3    & 4    &  5   &  6  &  7   \\
        \hline
        100 & &  3.72$~~~~$   &  4.30$~~~~$   &  4.67$~~~~$   &  5.58$~~~~$   & 5.37$~~~~$   &  6.18$~~~~$ &   & 3.59$~~~~$   &  4.46$~~~~$   &  4.96$~~~~$   &  5.87$~~~~$   &  6.27$~~~~$   &  7.40$~~~~$  \\
        500 & &  2.18   &  3.33   &  3.77   & 4.37   &  4.68   &  5.45 &   &  2.60   &  3.08   & 3.75   &  3.82   & 4.60   &  5.41  \\
        1000 & &  1.85   & 2.81   &3.27   &  3.69   & 4.36   & 4.88 &   & 1.71   & 2.48   &  2.69   & 3.04   &  3.93   & 4.45  \\
        2500 & &1.36   &  2.27   &  2.76   &  3.22   &  3.89   &  4.47 &   &  1.23   &  1.83   &  2.25   &  2.68   &  3.48   &  3.89  \\
        5000 & &  1.11   &  1.82   &  2.29   &  2.81   &  3.32   &  3.93 &   &  1.04   &  1.62   &  2.20   &  2.60   &  3.32   &  3.80  \\
        7500 & &  0.96   &  1.59   &  2.11   &  2.85   &  3.00   &  3.50 &   &  1.02   &  1.53   &  2.13   &  2.58   &  3.28   &  3.86  \\
        10000 & &  0.83   &  1.51   &  2.11   &  2.66   &  2.83   &  3.37 &   &  1.01   &  1.53   &  2.09   &  2.53   &  3.31   &  3.76  \\
        25000 & &  0.68   &  1.19   &  1.75   &  2.35   &  2.70   &  2.91 &   &  0.92   &  1.41   &  2.01   &  2.48   &  3.26   &  3.50  \\
        50000 & &  0.51   &  1.08   &  1.54   &  2.04   &  2.37   &  2.80 &   &  0.84   &  1.26   &  1.79   &  2.56   &  3.00   &  3.05  \\
        75000 & &  0.48   &  0.98   &  1.35   &  1.93   &  2.22   &  2.63 &   &  0.76   &  1.12   &  1.70   &  2.41   &  2.60   &  3.09  \\
        \hline
    \end{tabular}
    \label{tab:Dwisebreakup}
\end{table*}
For 11~k molecules
with 3--7 substitutions (see Table \ref{tab:enum}), randomly drawn 200 from each set, was 
added to the hold-out set amounting to 1,200 molecules. We benchmarked the performance of KRR-QML models using
hold-out errors across four different representations: 1-hot, BoB, FCHL and SLATM.

\begin{figure}[!htp]
    \includegraphics[width=8.5cm]{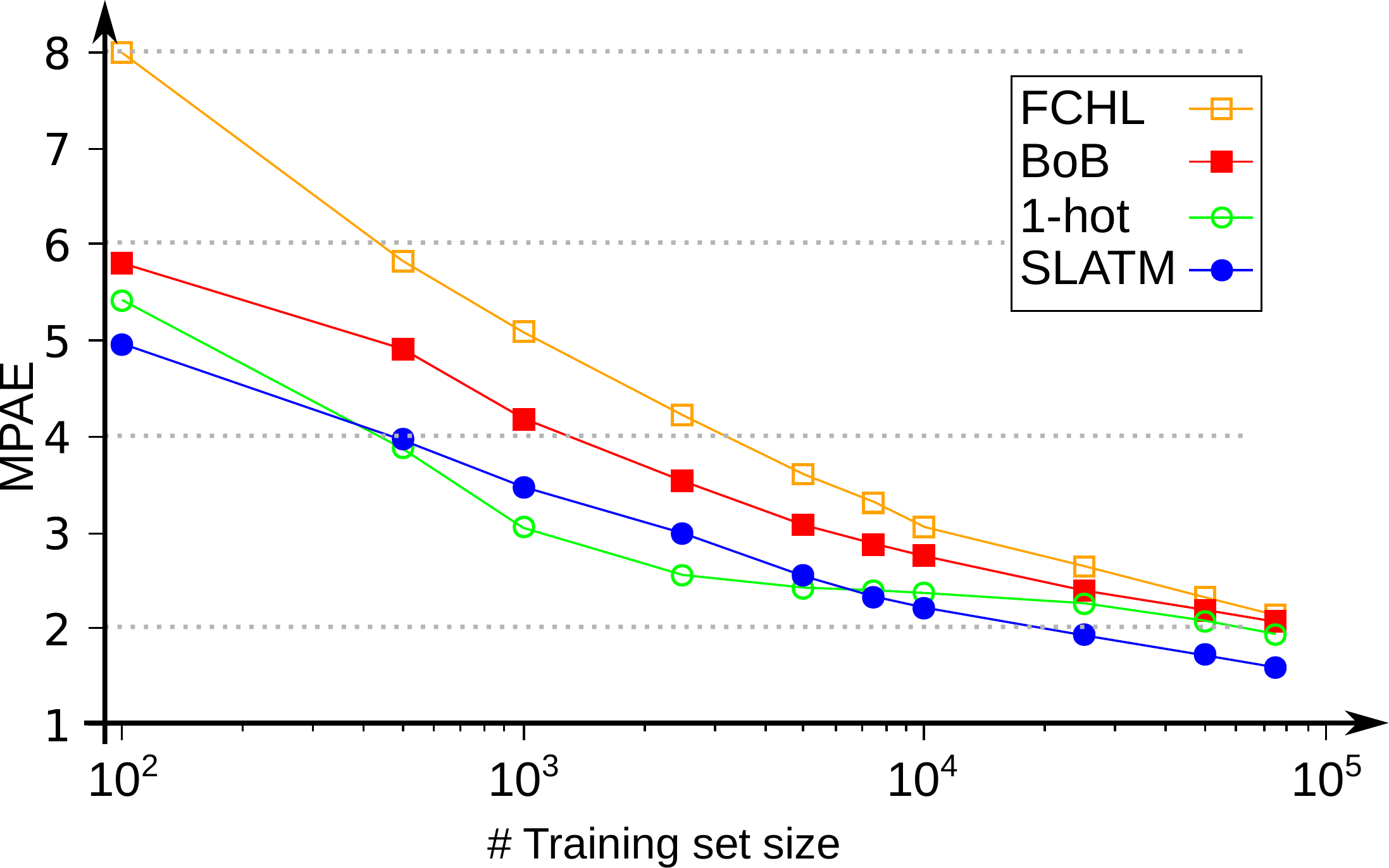}
    \caption{Learning rates for KRR-QML models based on
    the structure-based descriptors---FCHL, BoB, and SLATM---and a
    composition-based 1-hot representation. 
    Mean percentage absolute error (MPAE) for predicting TDDFT-level ${\rm S}_0\rightarrow{\rm S}_1$
    excitation energies of 1,200 hold-out BODIPY derivatives is shown for
    varying training set sizes.}
    \label{fig:slatmtraining}
\end{figure}

In FIG.~\ref{fig:slatmtraining}, we present the performance of QML models. 
When increasing the training set size to 75 k, all models show essentially monotonous convergence upon validating
on a 1,200 hold-out set. Of the four representations, SLATM shows the best performance 
at the 75~k limit with a mean percentage absolute error (MPAE) of 1.6\%. 
For a conventional dye such as Nile Red, with $\lambda_{max}$ corresponding to 2.41 eV\cite{zuehlsdorff2019modeling}, this MPAE translates to an uncertainty of $<0.05$ eV, which is well within the uncertainty of the target TDDFT method. 
The 1-hot, FCHL and BoB representations converge to MPAEs in the 1.9--2.1\% range. 
Compared to the structural representations, the remarkable accuracy of the
composition-based representation, 1-hot, may be ascribed to the significant influence of 
substituent type and site on the overall BODIPY excitation energies
rather than three-dimensional structural information. Similar observation has been made
in past works\cite{faber2016machine,ward2017including,kayastha2021machine,heinen2021toward}.
In the following, we use the best performing SLATM-KRR-QML model.

Although the MPAE for the 1,200 hold-out set indicates the QML models to provide accurate results in agreement with 
1-hot-KRR-QML models calculated separately for 200 entries from
each subset with 2--7 substitutions. In all categories, both representations converge to
small MPAEs with increasing model size, with
SLATM delivering better results. 
Both models provide least errors for doubly substituted derivatives while the worst results are noted for septuply substituted ones. The reason for the deterioration of the models' performance 
with increasing number of substituents is because highly substituted derivatives are under-represented in the training set. 
However, it must be noted for the most diverse case of septuply substituted BODIPYs, SLATM and 1-hot based models delivered
MPAEs of 2.63 and 3.09\% despite using only 0.0000296\% of the total space.
Hence, we conclude SLATM-KRR-QML and 1-hot-KRR-QML models to be rapid and accurate alternatives for first-principles high-throughput
modeling. Further, by exploiting the excellent QML cost to accuracy trade-off we demonstrate the
applicability of the 1-hot-KRR-QML model in the form of a publicly accessible web interface for navigating the
chemical space of BODIPYs with 253 giga molecules, see Appendix.


 \begin{figure*}[!htp]
    \includegraphics[width=0.85\linewidth]{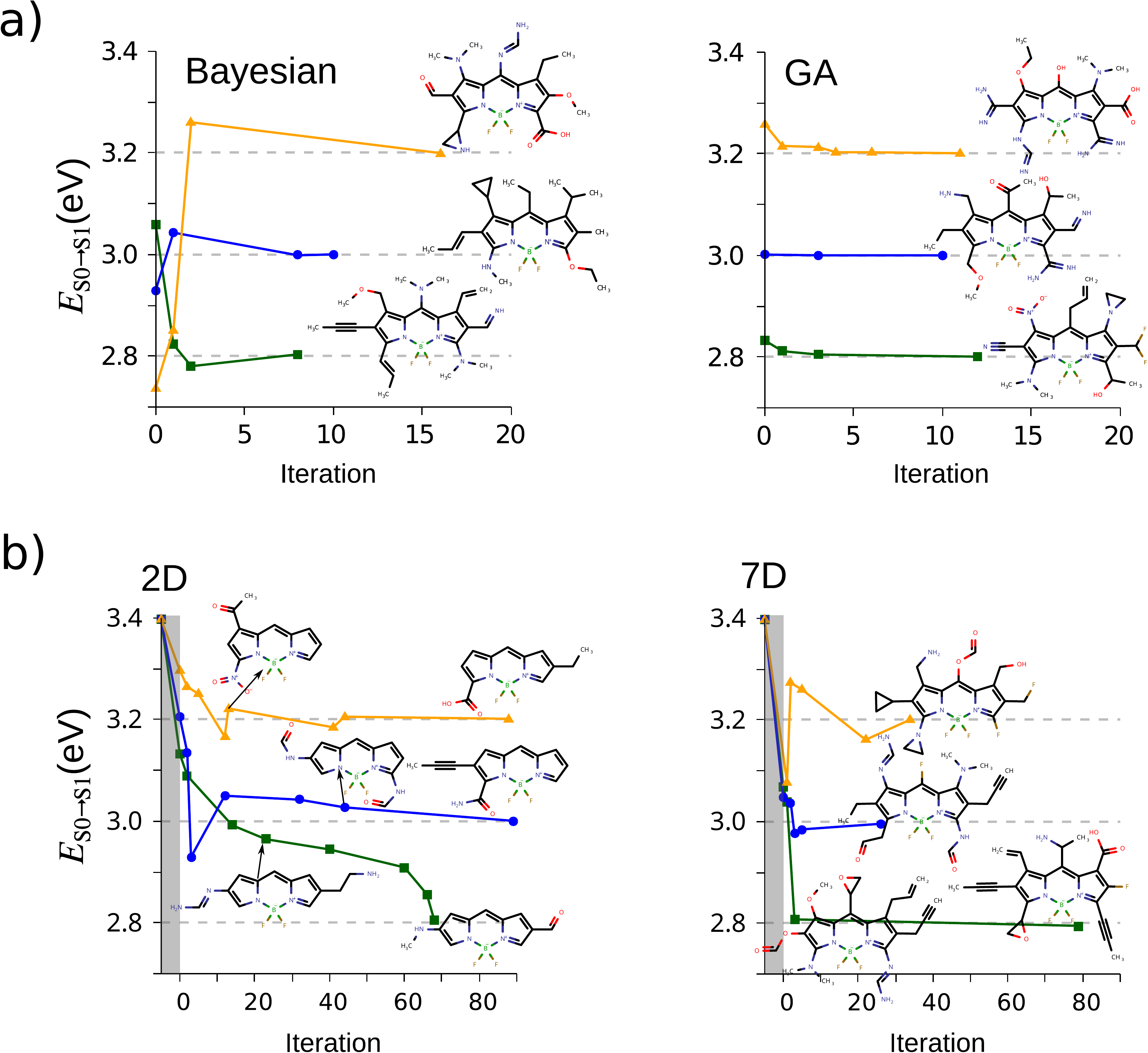}
    \caption{
    Convergence of inverse design searches of BODIPY molecules with ${\rm S}_0\rightarrow{\rm S}_1$ excitation 
    energy targets: 2.8 eV, 3.0 eV, and 3.2 eV:
    a) Comparison of unconstrained Bayesian 
    optimization and genetic algorithm (GA).  In each case, the final, optimal solutions are shown. 
    b) Bayesian search was performed separately in doubly substituted (2D) and septuply substituted (7D) subspaces.
    Gray region marks the first 5 iterations for building a Gaussian process model. Navigation trajectories start 
    from the unsubstituted BODIPY (3.40 eV). Intermediate solutions are also shown.
    The target property was calculated using a SLATM-KRR-QML model, trained on 
    TDDFT data for 75,000 examples. To accelerate predictions,
    PM7 level geometries were used for generating the SLATM descriptor. Python 
    codes for inverse design are maintained at \href{https://github.com/moldis-group/DesignBODIPY}{https://github.com/moldis-group/DesignBODIPY}.
    }
    \label{fig:Bayes}
 \end{figure*}

\subsection{Inverse designing BODIPY derivatives}
Inverse design offers a very economic solution
to zero-in on molecules with desirable properties because the solution is sought iteratively without having to 
screen through all possibilities in an Edisonian approach. 
Inverse design is a mathematically ill-posed problem due to a
surjective ({\it i.e.,} many-to-one) mapping between the chemical structure and the target property. However, when the target
property value is known to correspond to one or many solutions, state-of-the-art algorithms provide
optimal solutions with in a numerical precision. Here, we explore the possibility to inverse design BODIPY derivatives with fixed
${\rm S}_0\rightarrow{\rm S}_1$ excitation energy targets. The function values (property) required for the inverse design
optimizers can ideally come from TDDFT calculations albeit at a higher computational overhead. Hence, we use the SLATM-KRR-QML model to 
estimate the property because of favorable accuracy-vs-speed.

While it may be desired for inverse design to search for systems with extreme property values, its performance
is dependent on the knowledge included in the property generator. Our models were trained on a randomly drawn subset of 
the total chemical space. Most molecular properties result in peaked distributions with sparsely populated tails. Hence, molecules
at these extreme regions of the distributions will be under-represented in any randomly sampled training set. 
This suggests inverse design based on QML to be less reliable for identifying errors when the target property is in a region of space that is under-represented in the training set, namely the extremes. Hence, it is recommended to perform inverse design for targets belonging to regions in the property space that was adequately represented in the training set for a QML based property-generator.
Here, we investigate the applicability of QML guided inverse design via two commonly used optimization protocols---the Bayesian Optimization and Genetic Algorithm (GA).

In FIG.~\ref{fig:Bayes}a, we note the performance of unconstrained searches across the BODIPY chemical space. Our targets 2.8 eV, 3.0 eV, and 3.2 eV are all red-shifted compared to the unsubstituted BODIPY, and sufficiently represented in the training set. The prevalence of septuply substituted BODIPY molecules as targets could be expected, as it comprises the largest BODIPY sub-space (see Table.~\ref{tab:enum}). GA requires 20 seeds (not shown in FIG.~\ref{fig:Bayes}) to pre-condition the optimizer. Hence, for Bayesian optimization, we arrive at BODIPY molecules with target S$_0\rightarrow$S$_1$ values in fewer iterations than that in GA. Also in Bayesian optimization, only those iterations are considered for which the loss function, (target - predicted)$^2$, is greater than the value from the previous iteration. While all searches concluded in septuply substituted BODIPYs for both Bayesian search and GA, it is likely that there are derivatives with $<$ 7 substitutions satisfying the same design target. Hence, this warrants a constrained search in BODIPY sub-spaces where GA will be particularly challenging. Hence, the Bayesian approach is ideal for a constrained inverse design.
In the doubly substituted subspace, the candidate structure with a 
target property value of 2.8 eV, corresponds to a (2,5) derivative. This seem to be in accord with the trend that
these sites favor red-shifting. However, rationalization of the intermediate and final solutions in evolutionary searches
need not exhibit continuous or known trends because of the very nature of these inverse optimizations exploiting the surjective structure-property mapping\cite{hornby2006automated}.
Hence, in inverse design, an attempt to interpret the final optimal solutions is prone to \textit{post hoc fallacy}.







\section{Conclusions}
Quantum chemistry aided rational design of a dye molecule guides chemists to identify molecules with favorable excitation properties. Accurate descriptions of excited state characteristics calls for long-range corrected DFT level modeling or beyond. However, the traditional one molecule at-a-time paradigm becomes prohibitively expensive when navigating the BODIPY chemical space formed by systematic introduction of 46 small organic substituents at 7 sites through single bond connectivities, yielding $>253$ Billion molecules. 

In this study, we have enumerated the complete chemical space spanned by BODIPYs with various degrees of substitution. For statistical modeling, we sampled 77,412 derivatives from the entire chemical space. 
The resulting BODIPYs dataset contains, the unsubstituted molecule, all possible singly (184), and doubly substituted (22,287) derivatives along
with about 11,000 triply--septuptly substituted derivatives. In the subset comprising singly and doubly substituted derivatives, we identified site-specific chemical trends by screening. Since the BODIPY dyes are known to exhibit small Stokes shifts, the vertical excitation energies provided in this study can aid
experimental endeavours. For such attempts to be fruitful, 
the BODIPYs dataset should be enriched by 
incorporating systematic corrections through careful 
calibrations of the TDDFT results presented here 
using high-level wavefunction theories. 

We have presented evidences for the failure of an additivity model to estimate the shift in
BODIPY's excitation energy due to various substitutions. Hence, 
investigating the complete chemical space of BODIPY with $>$ 2 substituents presents a significant computational challenge.
To this end, we benchmarked the performance of KRR-QML models for inductive modeling of the lowest
excitation energy. Using DFT-level properties of 77 k example molecules for training the QML model, we compared
the performance of three structural
representations: SLATM, FCHL, and BoB, and a categorical 1-hot descriptor. 
QML model trained on 75 k BODIPY entries with the SLATM descriptor exhibits the 
best performance with an average error of $<2\%$ for a randomly drawn hold-out set. 
The 1-hot representation, that can be instantaneously generated, 
delivers the next best performance enabling the 
development of a publicly accessible web-based QML model enabling rapid and seamless query on the entire chemical space of BODIPY.
Using excitation energies predicted by a SLATM-KRR-QML model, we inverse designed BODIPYs with target property values. We tested Bayesian optimization and GA and found the former to outperform the latter. Furthermore, 
with in the chemical subspaces for a given number of substituents, constrained Bayesian optimization was
performed to identify BODIPY molecules exhibiting target excitation energy values. 

\section{Data Availability}
PM7, DFT and TDDFT level properties of 77,412 molecules used for training a QML model
are available at \href{https://moldis-group.github.io/BODIPYs}{https://moldis-group.github.io/BODIPYs}.
A QML model to predict S$_0$$\rightarrow$S$_1$ 
excitation energy of BODIPYs accessible via a web browser
is available at
\href{https://moldis.tifrh.res.in/db/bodipy}{https://moldis.tifrh.res.in/db/bodipy}.

\section{Acknowledgments}
We acknowledge support of the Department of Atomic Energy, Government
of India, under Project Identification No.~RTI~4007. 
All calculations have been performed using the Helios computer cluster, which is an integral part of the MolDis Big Data facility, TIFR Hyderabad \href{http://moldis.tifrh.res.in}{(http://moldis.tifrh.res.in)}.

\appendix*
\section{Navigating the BODIPY chemical space using a web-based QML model}
\begin{figure}[!htbp]
    \centering
    \includegraphics[width=\linewidth]{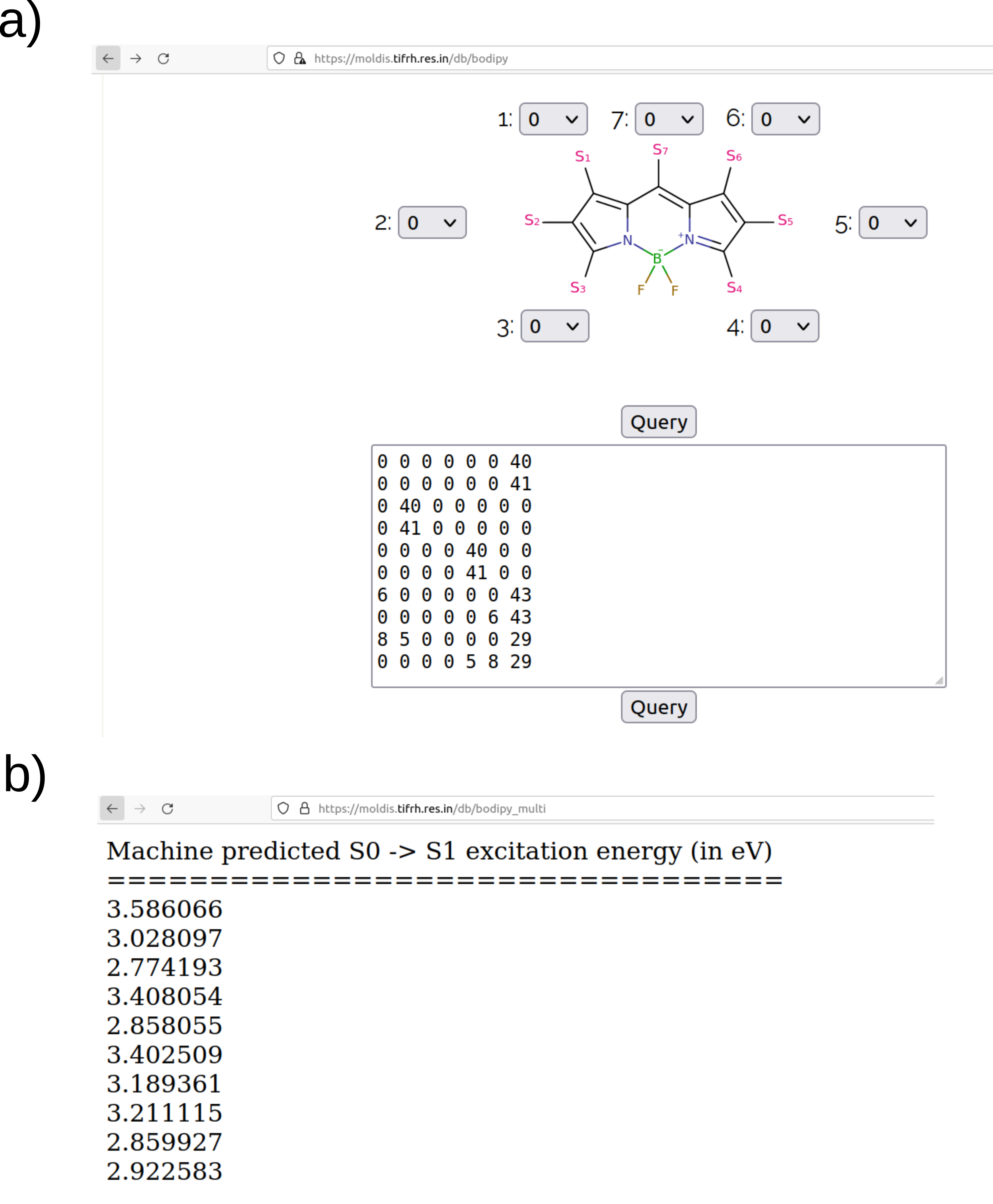}
    \caption{Screenshots of the web-based interface for QML predictions of ${\rm S}_0\rightarrow{\rm S}_1$ excitation energy 
    in the BODIPY chemical space: a) query page, b) results page. The interface can be accessed at
    \href{https://moldis.tifrh.res.in/db/bodipy}{https://moldis.tifrh.res.in/db/bodipy}.}
    \label{fig:webapp}
 \end{figure}

 We provide a web-based interface for a QML model using the 1-hot representation, trained on
 75 k BODIPY derivatives, see FIG.~\ref{fig:webapp}. The input in the interface can be provided in two ways:
 1) using drop-down menus one can select any of the 46 substituents at each of the 7 sites, or
 2) to query multiple molecules, one can manually list the indices of substituents at sites 1--7 in the same order. 
 The inputs in FIG.~\ref{fig:webapp}a illustrate some of the concepts discussed in the main text.
 Influence of electron donating/withdrawing groups at site-7 is opposite to that of hosting the same groups at site-2 or site-5.
 For BODIPY, the ${\rm S}_0\rightarrow{\rm S}_1$ excitation energy is at 3.40 eV. Compared to this, the single substituted
 derivative with group-40 at site-7 is blue-shifted to 3.59 eV, see FIG.~\ref{fig:webapp}b. While the value for group-41 at the same site is
 red-shifted to 3.03 eV (FIG.~\ref{fig:webapp}b). The red/blue shifting is reversed when these groups are at site-2. Even though site-5 is
 equivalent to site-2, one notices small deviations when comparing the results for same groups at both sites. This is because the QML
 model is trained on unique derivatives, hence the accuracy is differently influenced for chemically equivalent combinations. 

\bibliographystyle{apsrev4-1}
\bibliography{biblio.bib}
\end{document}